\documentclass[twocolumn,tighten,numberedappendix]{aastex63}
\usepackage{amsmath}

 \usepackage{lineno}
\def\simlt{\lower.5ex\hbox{$\; \buildrel < \over \sim \;$}}
\def\simgt{\lower.5ex\hbox{$\; \buildrel > \over \sim \;$}}

\def\beq{\begin{equation}}
\def\eeq{\end{equation}}
\def\ba{\begin{eqnarray}}
\def\ea{\end{eqnarray}}
\def\bB{\boldsymbol{B}}
\def\bE{\boldsymbol{E}}
\def\bS{\boldsymbol{S}}

\def\Sect{{\rm Section}}

\def\Eq{Eq.}
\def\Eqs{Eqs.}

\def\M{{\cal M}}

\def\sT{\sigma_{\rm T}}

\def\E{{\cal E}}
\def\N{{\cal N}}

\def\RLC{R_{\rm LC}}

\def\omL{\omega_{\rm L}}

\def\vgr{v_{\rm gr}}
\def\ggr{\gamma_{\rm gr}}

\def\omB{\omega_B}

\def\Bbg{B_{\rm bg}}

\def\bBbg{\bB_{\rm bg}}
\def\bEbg{\bE_{\rm bg}}
\def\Bb{\hat{B}_{\rm bg}}
\def\bBb{\hat{\bB}_{\rm bg}}
\def\Eb{\hat{E}_{\rm bg}}
\def\bEb{\hat{\bE}_{\rm bg}}
\def\omBb{\hat{\omega}_B}

\def\vph{v_{\rm ph}}

\def\sgg{\sigma_{\gamma\gamma}}

\def\gp{\gamma_p}

\def\vp{v_p}
\def\bp{\beta_p}

\def\bbg{\beta_{\rm bg}}

\def\omsc{\omega_c}

\def\ssc{\sigma_{\rm sc}}

\def\epsc{\epsilon_{c}}

\def\dE{\dot{\E}}

\def\gL{\gamma_{\rm RRL}}

\def\N{{\cal N}}

\def\as{a_\star}

\def\Ra{R_0}
\def\Esc{{\cal E}_{\rm sc}}

\def\tausc{\tau_{\rm sc}}
\def\tcross{t_{\rm cross}}

\def\nco{n_{\rm co}}

\newbox\grsign \setbox\grsign=\hbox{$>$} \newdimen\grdimen \grdimen=\ht\grsign
\newbox\simlessbox \newbox\simgreatbox \newbox\simpropbox
\setbox\simgreatbox=\hbox{\raise.5ex\hbox{$>$}\llap
     {\lower.5ex\hbox{$\sim$}}}\ht1=\grdimen\dp1=0pt
\setbox\simlessbox=\hbox{\raise.5ex\hbox{$<$}\llap
     {\lower.5ex\hbox{$\sim$}}}\ht2=\grdimen\dp2=0pt
\setbox\simpropbox=\hbox{\raise.5ex\hbox{$\propto$}\llap
     {\lower.5ex\hbox{$\sim$}}}\ht2=\grdimen\dp2=0pt
\def\simgt{\mathrel{\copy\simgreatbox}}
\def\simlt{\mathrel{\copy\simlessbox}}

\defcitealias{Beloborodov21}{B21}

\begin{document}

\title{Can a strong radio burst escape the magnetosphere of a magnetar?}

\author{Andrei M. Beloborodov}

\affiliation{Physics Department and Columbia Astrophysics Laboratory, Columbia University, 538 West 120th Street, New York, NY 10027, USA}

\affiliation{Max Planck Institute for Astrophysics, Karl-Schwarzschild-Str. 1, D-85741, Garching, Germany}

\begin{abstract}
We examine the possibility that fast radio bursts (FRBs) are emitted inside the magnetosphere of a magnetar. On its way out, the radio wave must interact with a low-density $e^\pm$ plasma in the outer magnetosphere at radii $R=10^9$-$10^{10}\,$cm. In this region, the magnetospheric particles have a huge cross section for scattering the wave. As a result, the wave strongly interacts with the magnetosphere and compresses it, depositing the FRB energy into the compressed field and the scattered radiation. The scattered spectrum extends to the $\gamma$-ray band and triggers $e^\pm$ avalanche, further boosting the opacity. These processes choke FRBs, disfavoring scenarios with a radio source confined at $R\ll 10^{10}\,$cm. Observed FRBs can be emitted by magnetospheric flare ejecta transporting energy to large radii.
\end{abstract}


% \maketitle

%#####################################################################

 \section{Introduction}

Fast radio bursts (FRBs) are a big puzzle. They are detected from cosmological distances with luminosities up to $10^{43}$erg/s and ms durations \citep{Petroff19}. Spectacular progress in FRB observations, in particular by CHIME, has provided a wealth of data which miss theoretical explanation. 

Recent FRB detection from SGR~1935+2154 \citep{CHIME20,Bochenek20} supports the association of FRBs with magnetars. However, their emission mechanism is not established. Possible scenarios are (1) a radio source confined near the magnetar, inside its ultrastrong magnetosphere (``near-field'') and (2) emission at much larger radii from explosions launched by magnetospheric flares into the magnetar wind (``far-field''). The far-field models include a robust radiative mechanism---synchrotron maser emission from the blast wave \citep{Lyubarsky14,Beloborodov17b,Beloborodov20,Metzger19,Sironi21} and possible emission from the magnetic flare ejecta \citep{Lyubarsky20}. On the other hand, it has been argued that the complex temporal structure detected in FRBs (assuming it forms inside the source, not via propagation effects) favors the near-field scenario \citep{CHIME21,Nimmo21}. 

Cosmological FRBs are $10^{10}$-$10^{12}$ times brighter than radio pulsations detected in several magnetars \citep{Kaspi17}, and proposals for magnetospheric radio emission mechanisms \citep{Thompson08,Beloborodov13a,Lyutikov16,Lu20} face challenges when applied to FRBs \citep{Lyubarsky21}. In this {\it Letter}, we do not rely on any concrete emission mechanism, and instead examine a generic constraint. 

A simple and essential requirement for any near-field FRB scenario is the successful escape of the radio wave. The emitted wave must propagate through the plasma in the closed magnetosphere, which extends to the light cylinder
\beq
   \RLC=\frac{c}{\Omega}\approx 5\times 10^{9}\,\left(\frac{P}{1\,\rm s}\right) {\rm cm},
\eeq
where $P=2\pi/\Omega$ is the rotation period of the magnetar. The plasma is immersed in the magnetospheric field $\Bbg$ and has gyro-frequency
\beq
   \omega_B=\frac{e\Bbg}{m_ec}\approx \frac{e\mu}{m_ecR^3}\approx 1.8\times 10^{13}\,\frac{\mu_{33}}{R_9^3}\,{\rm rad/s},
\eeq
where $e$ and $m_e$ are the electron charge and mass, and the magnetic dipole moment  $\mu\approx \Bbg R^3\sim 10^{33}\,$G\,cm$^3$ corresponds to a surface magnetic field $B_\star\sim 10^{15}\,$G. The radio wave frequency $\nu=\omega/2\pi\sim 1$~GHz satisfies $\omega<\omB$ at radii up to $\RLC$  if $P<3\,\mu_{33}^{1/2}\,$s.
 
The magnetosphere is normally assumed to allow free escape of radio waves polarized perpendicular to the background magnetic field $\bBbg$, especially for $\omega\ll\omB$.  A standard calculation of the electron cross section for wave scattering gives $\ssc\approx (\omega/\omega_B)^2\sT\ll \sT$ \citep{Canuto71} where $\sT$ is the Thomson cross section. Recent discussions of FRB interaction with the magnetosphere  \citep{Lyutikov20,Kumar20} also concluded that the interaction is suppressed by $\omB\gg\omega$. 

However, the small $\ssc$ holds only for wave amplitudes $E_0<\Bbg$. In fact, as the wave propagates away from the magnetar in the decreasing background $\Bbg\propto R^{-3}$, its amplitude $E_0\propto R^{-1}$ eventually exceeds $\Bbg$:
\beq
\label{eq:Ra}
   \frac{E_0}{\Bbg} =\left(\frac{R}{\Ra}\right)^2,
   \qquad \Ra\approx 3.5\times 10^8 \,\frac{\mu_{33}^{1/2}}{L_{42}^{1/4}}{\rm ~cm},
\eeq
where $L=cR^2 E_0^2/2$ is the isotropic equivalent of the FRB luminosity.  
Hereafter we focus on the wave propagation at radii $R>\Ra$ where $E_0>\Bbg$. This condition is equivalent to $a_0>\omega_B/\omega$, where 
\beq
a_0=\frac{eE_0}{m_ec\,\omega}
     \approx 2.3\times 10^4 \,\frac{L_{\rm 42}^{1/2}}{R_9\,\nu_9}.
\eeq

Waves with $E_0>\Bbg$ are scattered with $\ssc$ exceeding $\sT$  by many orders of magnitude \citep[][hereafter B21]{Beloborodov21}. This occurs because particles ($e^\pm$) exposed to the wave are quickly accelerated to huge Lorentz factors, exceeding $10^4$, and emit $\gamma$-rays at the expense of the radio wave energy. Below we examine implications of these processes for FRBs. For definiteness we consider a plane wave propagating perpendicular to $\bBbg$; for different propagation angles $\theta\neq \pi/2$ one should replace $\omB\rightarrow\omB\sin\theta$.

%#####################################################

\section{Scattering of the wave}

\subsection{Magnetospheric plasma density $n_0$} 
\label{n0}

The scattering opacity of the magnetosphere depends on its initial plasma density  $n_0$ before it is exposed to the outgoing ultrastrong wave. One characteristic density in the problem is $\nco=|\nabla\cdot\bE_{\rm co}|/4\pi e$, the minimum needed to sustain the co-rotation electric field $\bE_{\rm co}=\bBbg\times(\boldsymbol{\Omega}\times\boldsymbol{R})$ \citep{Goldreich69}, 
\beq
\label{eq:nco}
  n_{\rm co}  \sim \frac{\mu}{ceR^3P}\approx 
   7\times 10^4 \,\frac{\mu_{33}}{R_9^3}
   \left(\frac{P}{1\,\rm s}\right)^{-1} {\rm cm}^{-3}.
\eeq 
As shown below, even this low density is sufficient to choke FRBs emitted at $R\ll \RLC$.

The actual density around active magnetars is higher than $\nco$. The $e^\pm$ plasma is created near the neutron star, at a radius $R_\pm\simlt 10 R_\star$, and fills the outer magnetosphere by flowing out along the extended magnetic field lines. The particle flow $\dot{\N}$ at a radius $R\gg R_\pm$ scales with the number of field lines connecting $R_\pm$ and $R$, which implies $\dot{\N}(R)\propto R^{-1}$ \citep{Beloborodov20}.

The value of $\dot{\N}$ may be estimated using observations of known magnetars in our galaxy. Their typical persistent luminosity in the keV band is $L_{\rm keV}\sim 10^{35}\,$erg/s, and their spectra are nonthermal, extending with a hard slope to energies $\gg 10\,$keV \citep{Kaspi17}. The spectrum is emitted by the magnetospheric $e^\pm$ flow, which radiates its energy away via resonant cyclotron scattering \citep{Thompson02,Beloborodov13a,Beloborodov13b}. The flow decelerates to a mildly relativistic speed at radius $R_{\rm keV} \approx 2\times 10^7\mu_{33}^{1/3}\,$cm (where $\hbar\omB\sim 1\,$keV), and the $e^\pm$ supply to $R_{\rm keV}$ may be roughly estimated as $\dot{\N}(R_{\rm keV})\sim L_{\rm keV}/m_ec^2$. This estimate is also consistent with the theoretically expected multiplicity of $e^\pm$ production around magnetars \citep{Beloborodov13b}.

Then, using the scaling $\dot{\N}(R)\propto R^{-1}$, one can estimate the $e^\pm$ density in the outer magnetosphere as
\beq
\label{eq:n0}
  n_0(R)\sim \frac{\dot{\N}(R)}{cR^2}\sim 10^{10}\,R_9^{-3} \left(\frac{L_{\rm keV}}{10^{35}\,\rm erg/s}\right) {\rm~cm}^{-3}.
\eeq
Note that \Eqs~(\ref{eq:nco}) and (\ref{eq:n0}) both give a density $\propto R^{-3}$. Therefore, a convenient density parameter is 
\beq
  \N \equiv n_0 R^3.
\eeq
\Eqs~(\ref{eq:nco}) and (\ref{eq:n0}) give $\N\sim 10^{32}$ and $10^{37}$, respectively.

\subsection{Bulk acceleration and compression}
\label{bulk}

 Before discussing the scattering opacity of the plasma, it is useful to look at the  response of the magnetosphere to a force exerted by the wave. The basic response is the sudden bulk acceleration and compression. 

The magnetic field $\Bbg$ plays a key role in the wave-plasma interaction (even when $\Bbg\ll E_0$). Without $\Bbg$, an initially static $e^\pm$ plasma exposed to the ultrastrong radio burst would immediately develop outward bulk motion with a huge Lorentz factor $\gp\approx a_0$ (B21). Most of the radio wave would not even interact with the plasma, as the plasma would surf the leading front of the wave, unable to penetrate it. $\Bbg\neq 0$ plays a key role by arresting the ultra-relativistic bulk motion and so enforcing wave-plasma interaction. Larmor rotation of particles in $\Bbg$ effectively couples the plasma to the background magnetic field lines. On scales larger than the Larmor scale, the plasma behaves as an MHD fluid. This fluid still experiences some acceleration as it interacts with the wave, but now the bulk acceleration is controlled by the effective inertial mass of the magnetic field, $\Bbg^2/4\pi c^2$, which exceeds the plasma rest mass density $n_0m_e$ by many orders of magnitude.

The cross section for wave scattering will be defined in the fluid rest frame. We now evaluate the speed of this frame (plasma bulk speed $\beta_p=\vp/c$) assuming $\bBbg$ perpendicular to the wave propagation direction ($\theta=\pi/2$). Extension to oblique $\bBbg$ is straightforward: one should make an additional Lorentz boost along $\bBbg$ such that the wave propagation becomes perpendicular to $\bBbg$ in the boosted frame.

The MHD coupling implies that the momentum received by the plasma from the wave  is shared with the background field, so there appears an outward Poynting flux $\bS_{\rm bg}=(c/4\pi)\bEb\times\bBb$ and the magnetic field lines begin to drift with speed $\bbg=\Eb/\Bb$. Here, $\bEb$ and $\bBb$ are the new values of the background field, changed from the pre-wave values $\bBbg$ and $\bEbg=0$. In the ideal MHD approximation,  the plasma and the background field drift with equal speed,
\beq
   \bp=\bbg=\frac{\Eb}{\Bb}.
\eeq
Acceleration to $\bp$ is accompanied by compression,
\beq
    \frac{\Bb}{\Bbg}=\frac{n}{n_0}=\frac{1}{1-\bp}.
\eeq 
The wave compresses the magnetosphere of radius $R$ into a shell of thickness $(1-\bp)R$.

Let $\E$ be the isotropic equivalent of the FRB energy and $\tausc<1$ be the scattering optical depth encountered by the radio wave as it propagates out to a radius $R$. Momentum lost by the wave $\tausc \E/c$ is mostly taken by the magnetospheric field, which dominates over plasma,
\beq
   (1-\bp)R\, \frac{S_{\rm bg}}{c^2}\approx \frac{\tausc\E}{4\pi R^2c},
\eeq
where $S_{\rm bg}/c^2=\bp\Bb^2/4\pi c$ is the momentum density of the background field. This gives the equation for $\bp$,
\beq
\label{eq:bp}
   \frac{\bp}{1-\bp}\approx \frac{\tausc\E}{R^3\Bbg^2}\approx \tausc\,R_9^3\,\frac{\E_{39}}{ \mu_{33}^2}.
\eeq
One can see that any significant scattering at $R>10^9\,$cm is accompanied by sweeping the outer magnetosphere to a high Lorentz factor $\gp=(1-\bp^2)^{-1/2}$. This bulk acceleration must be self-consistently taken into account when calculating the wave scattering.

The wave has a small duration $T\sim 1\,{\rm ms}\ll R/c$. The compressed magnetospheric shell crosses the wave and exits behind it on the timescale $\tcross=T/(1-\bp)$. As long as $\tcross<R/c$, the plasma is not stuck in a leading portion of the wave, and the entire wave interacts with the plasma. The condition $\tcross<R/c$ is satisfied at 
\beq
   R<R_{\rm surf}\approx 6\times 10^9\,\tausc^{-1/2}\,\E_{39}^{-1/2}T_{-3}^{-1/2} \mu_{33} {\rm ~cm}.
\eeq
For typical parameters, $R_{\rm surf}$ is comparable to $\RLC$, and $\tcross<R/c$ is satisfied in the entire magnetosphere.

\subsection{Scattering optical depth}

For a successfully escaping wave, the net energy lost to scattering $\Esc$ must be small compared with the wave energy $\E=4\pi R^2 F T$, where $F=cE_0^2/8\pi$ is its energy flux. The lost energy fraction may be written as
\beq
   \tau_{\rm sc}=\frac{\Esc}{\E}  \sim n_0 R\, \ssc',   \qquad \E=LT,
\eeq
where $n_0$ is the initial (pre-wave) plasma density, and $\ssc'$ is the scattering cross section of $e^\pm$ measured in frame $K'$ that moves with speed $\bp$ (the rest frame of the accelerated and compressed magnetospheric shell).

Hereafter all quantities with primes  refer to frame $K'$. In this frame, the background field is purely magnetic: $\bBbg'\neq 0$, $\bEbg'=0$. Transformations of the wave frequency $\omega$ and field $\Bb$ to frame $K'$ give
\beq
   \omega'=\frac{\omega}{\gp(1+\bp)}, \qquad \omB'=\frac{\omBb}{\gp}
   =\gp(1+\bp)\omB.
\eeq
Wave strength parameter $a_0'=a_0$ is Lorentz invariant.

In frame $K'$, a minimum Lorentz factor developed by particles oscillating in the wave is $\gamma'\sim a_0$.\footnote{$\omB'>\omega'$ does not prevent particle acceleration by $E_0'\gg\Bbg'$. The wave accelerates the particle to a high $\gamma'$ and reduces its Larmor frequency $\omL'=\omega_B'/\gamma'$ before the particle has a chance to complete one gyration in $\Bbg'$ (B21).
}
Each particle emits curvature radiation with power $\dE_e'=\dE_e$, which determines the effective scattering cross section $\ssc'=\dE_e/F'$. To get an idea of the importance of scattering, one can start with a simple guess that particles in waves with $E_0'\gg \Bbg'$ emit  similarly to the known result at $\Bbg'=0$: $\dE_e\sim e^2\omega'^2\gamma'^4\sim e^2\omega'^2 a_0^4$, which gives $\ssc'\sim a_0^2\sT$ \citep{Landau75}. Then, one finds 
\beq
   \tausc\sim\frac{\ssc'\N}{R^2}
   \sim \frac{3\times 10^{3}}{R_9^4}\,\frac{L_{\rm 42}}{\nu_9^2}
   \left(\frac{\ssc'}{a_0^2\sT}\right) \N_{37}.
\eeq
The actual effect of $\Bbg'\ll E_0'$ on $\ssc'$ is not small and can give $\ssc'\gg a_0^2\sT$ (\Sect~\ref{ssc}). Furthermore, we will show below that even a much smaller $n_0\sim\nco$ ($\N\sim 10^{32}$) provides sufficient seeds for $e^\pm$ avalanche, which again leads to $\tausc\gg 1$.

\subsection{Scattering cross section}
\label{ssc}

Cross section $\ssc'=\dE_e/F'$ expresses irreversible energy losses of the radio wave, caused by the radiative power $\dE_e$ of each particle oscillating in it. The produced radiation has a huge characteristic frequency $\omsc'\approx a_0\gamma'^2\omega'$ (B21) and a small radiation formation length $l_f'\sim \gamma'^2c/\omsc'\sim (c/\omega'a_0)$. The inter-particle distance $l' \sim n'^{-1/3}\approx 10^{-3} R_9 (\gp\N_{36})^{-1/3}$\,cm is smaller than $c/\omega'$, but far greater than $c/\omsc'$. The radiative process is not coherent --- each particle radiates $\gamma$-rays independently (besides, the oscillation of the particles becomes incoherent when their orbits develop chaos, as described in B21). Therefore, $\ssc'$ can be found by examining the behavior of a single particle in the wave. 

One can also verify that the plasma is unable to screen the radio wave, as the maximum electric current $j_{\max}'=cen'$ is  far below the wave displacement current $\omega' E_0'/4\pi$. Their ratio equals $(\omega_p'/\omega')^2(\gamma'/a_0)$, where 
\beq 
   \omega_p'=\left(\frac{4\pi e^2 n'}{m_e\gamma'}\right)^{1/2}
\eeq
 is the plasma frequency and $n'=n/\gp=\gp(1+\bp)n_0$ is the particle density in frame $K'$. Any collective processes on the plasma timescale $\omega_p'^{-1}$ are slow and negligible compared to wave scattering by individual particles. This is an interesting  special feature of strong waves. 

%%%%%%%%%%% FIGURE %%%%%%%%%%%%%%%%%%
\begin{figure}[t]
\includegraphics[width=0.45\textwidth]{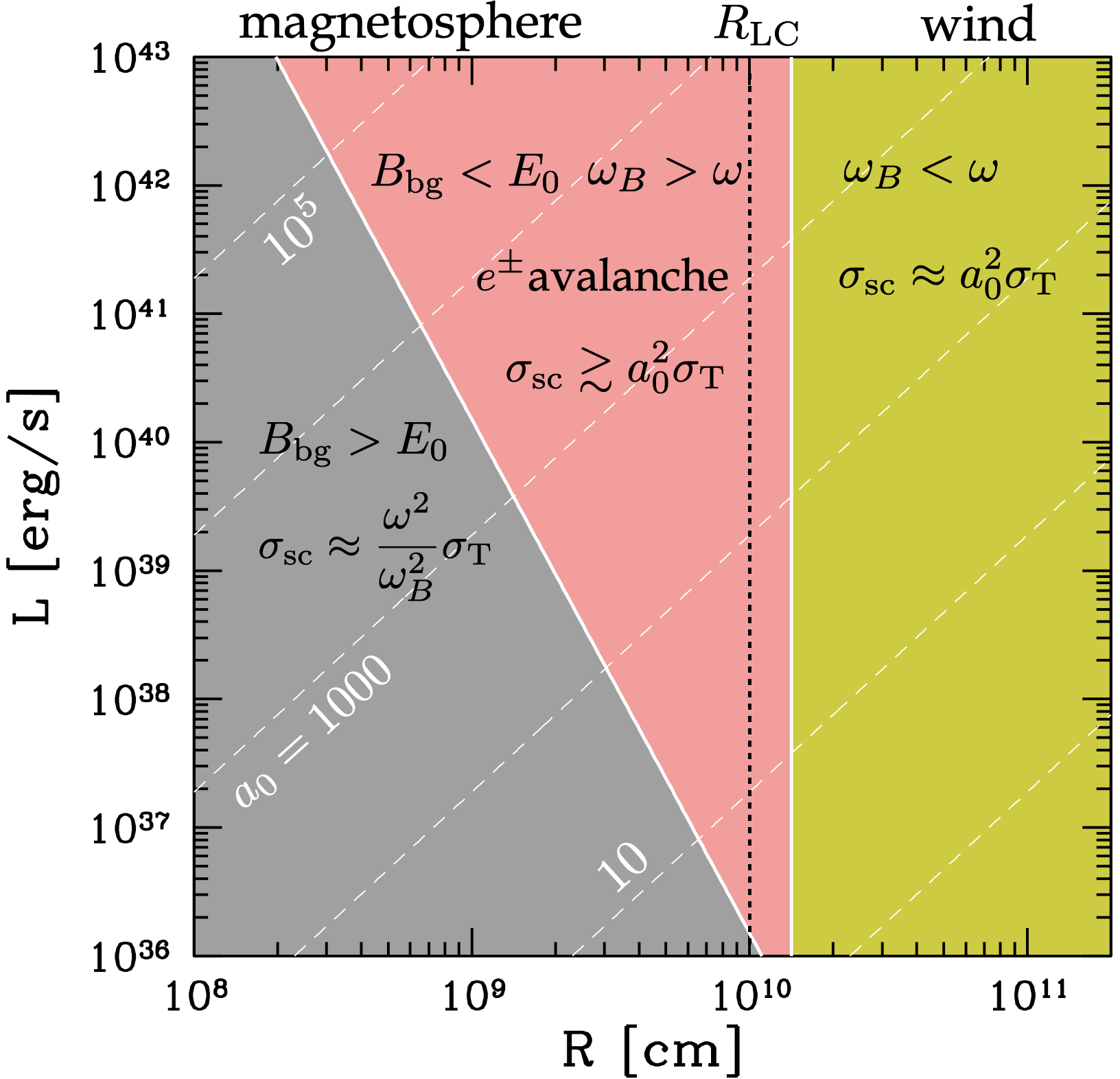} 
\caption{Three regions on the radius-luminosity plane for an FRB with $\nu=\omega/2\pi=1\,$GHz. The magnetar is assumed to have a magnetic dipole moment $\mu=10^{33}\,$G\,cm$^3$ and spin period $P=2\,$s. In the grey region $\Bbg>E_0$ ($R<R_0$), scattering is suppressed. In the region of $1<\omB/\omega<a_0$ (pink), strong scattering and the avalanche of $e^\pm$ creation occur, which choke the FRB. In the region of $\omB<\omega$ (olive), the scattering opacity quickly falls off. The $\omB=\omega$ boundary between  scattering regimes I (olive) and II (pink) is shown in the limit of $n_0\rightarrow 0$ ($\bp=0$, $\ssc'=\ssc$). With increasing $\tausc$, the plasma bulk acceleration expands the pink region.
 }
\label{fig:LR}
 \end{figure}
%%%%%%%%%%% FIGURE %%%%%%%%%%%%%%%%%%

B21 calculated the particle motion in frame $K'$ and found $\ssc'$ for two regimes (which approximately correspond to the olive and pink regions in Fig.~\ref{fig:LR}):

(I) If $\omB'<\omega'$, the particle motion is dominated by the $\omega'$-oscillations in the wave. Then,
\beq
\label{eq:ssc1}
  \frac{\ssc'}{a_0^2\sT}\approx \left\{\begin{array}{ll}
           1 & \quad a_0<a_1' \\
           (a_0/a_1')^{-0.6} & \quad a_0>a_1'
                                        \end{array}\right. 
\eeq
\beq
   a_1'\equiv \left(\frac{c}{r_e\omega'}\right)^{1/4}\sim 3\times 10^3, \qquad r_e\equiv\frac{e^2}{m_ec^2}.
\eeq

(II) If $\omB'>\omega'$ and the wave has strength $a_0<\as'\approx(c\,\omB'^2/r_e\omega'^3)^{1/5}$, the particle motion becomes simple Larmor rotation in $\Bbg'$ with frequency $\omL'=\omB'/\gamma'<\omega'$ (with superimposed subdominant $\omega'$-oscillations in the wave). The Larmor rotation is pumped because the particle experiences a resonance  with the wave every Larmor period, as shown in detail in B21. As a result, the particle's Lorentz factor is  quickly pushed to the  radiation reaction limit (RRL),
\beq
\label{eq:gL}
  \gL'\approx \left(\frac{c}{r_e\omega' a_0}\right)^{3/8}\left(\frac{\omB'}{\omega'}\right)^{1/4},
\eeq
and the resulting scattering cross section is 
\beq
\label{eq:ssc}
  \frac{\ssc'}{\sT}\sim \gL'^2 
  \approx 2\times 10^8 \,\frac{[\gp(1+\bp)]^{7/4} \mu_{33}^{1/2}}{R_9^{3/4}L_{42}^{3/8}\nu_9^{1/2}}.
\eeq
The condition $a_0<\as'$ implies $\ssc'\gg a_0^2\sT$.  It is satisfied for FRBs with $L\simlt 10^{41}\gp^2$~erg/s, as seen from the relation
\beq
\label{eq:as}
    \frac{\as'}{a_0}
    \approx \frac{\gp(1+\bp)}{2.3}\,\frac{\mu_{33}^{2/5}\nu_9^{2/5}}{R_9^{1/5}L_{42}^{1/2}}.
\eeq
 
These results for $\ssc'$ hold for a wave packet with a characteristic frequency $\omega$ (numerical examples in B21 used a modulated sine wave). Note also that the approximate \Eqs~(\ref{eq:ssc1}) and (\ref{eq:ssc}) do not match at $\omB'=\omega'$ --- there is a sharp change of $\ssc'$ across this transition.

The above expressions for $\ssc'$ assume that the radio wave propagates with vacuum speed $c$. Deviations from $c$ are small, as follows from the dispersion relation $\omega'^2=c^2k'^2+\omega_p'^2$, where  $k'$ is the wavenumber of a Fourier mode.\footnote{Density $n_0'$ and the corresponding $\omega_p'$ are not modulated by Larmor rotation in $\Bbg$ even in regime~II, where particle motion is dominated by gyration with $\omL'\ll \omega'$. The plasma accelerated in the wave initially forms a thin stream in the phase space, but it quickly broadens as particles develop chaos in $\gamma'$ (B21).}
Phase speed $v_{\rm ph}'=\omega'/k'>c$ and group speed $v_{\rm gr}'=d\omega'/dk'<c$ are related by $v_{\rm ph}'v_{\rm gr}'=c^2$, and their deviations from $c$ are small if $\ggr'\equiv (1-\vgr'^2/c^2)^{-1/2} \gg 1$. In particular, in regime~II we find (using $\gamma'\sim\gL'$):
\beq
   \ggr'\approx\frac{\omega'}{\omega_p'}\sim \frac{5\times 10^3\,\mu_{33}^{3/8} \nu_9^{5/8}}{\tausc^{1/2}\,[\gp(1+\bp)]^{3/16} R_9^{1/16} L_{42}^{9/32}},
\eeq
where we substituted $n_0=\tausc/R\ssc'$ and $\ssc'\sim\gL'^2\sT$. The huge $\ggr'$ at $\tausc=1$ shows that scattering of the wave is a much stronger effect than dispersion. (Strong dispersion of the wave packet would require time $\sim \ggr'^2T'$, far exceeding the burst age in frame $K'$, $t'\sim R/c\gp$.) It also justifies the use of $\vph'=c$ in the calculation of $\ssc'$.

\section{Avalanche of pair creation}
\label{avalanche}

A magnetospheric particle exposed to the strong radio wave radiates the power $\dE_e=\ssc' F'$ in photons with characteristic frequency $\omsc'= (3/2)\gamma'^3 c/r_c'$, where $r_c'^{-1}=(3\dE_e/2ce^2\gamma'^4)^{1/2}$ is the curvature of the particle's trajectory. Using the results of B21, we find in regime~I:
\beq
   \epsc'\equiv \frac{\hbar\omsc'}{m_ec^2}\approx \,a_0^3\,\frac{\hbar\omega'}{m_ec^2}
   \approx \frac{100}{\gp(1+\bp)}
    \frac{L_{42}^{3/2}}{R_9^3\,\nu_9^2},
\eeq 
and in regime~II:
\beq
    \epsc'\approx  \frac{1}{\alpha} \left(\frac{r_e\omB'^2 a_0}{c\,\omega'}\right)^{1/4}
    \approx 44 \,\frac{[\gp(1+\bp)]^{3/4} \mu_{33}^{1/2} L_{42}^{1/8}}{R_9^{7/4}\nu_9^{1/2}},
\eeq
where $\alpha=e^2/\hbar c$. We conclude that the wave generates $\gamma$-rays with $\epsc'\gg 1$. Their spectrum has a peak with a half-width extending from $0.01\epsc'$ to $1.5\epsc'$ \citep[e.g.][]{Longair11}. This broad peak extends over the MeV band where photon-photon collisions $\gamma+\gamma\rightarrow e^++e^-$ occur with cross section $\sgg'\sim 0.1\sT$.

It is convenient to view $e^\pm$ creation in frame $K'$, where the emitted $\gamma$-rays are quasi-isotropic.\footnote{In regime~I with $a_0>a_1'$, the emission has a backward beaming in frame $K'$; however this regime is not relevant for most FRBs.}
The energy emitted as the wave expands to radius $R$, $\Esc'=\Esc/\gp$,  occupies radial thickness $\Delta R'\sim R/\gp$, so the $\gamma$-rays have energy density
\beq
   U_\gamma' \approx \frac{\tausc\E}{4\pi R^3}.
\eeq
 They exit behind the wave on a timescale comparable to the wave duration $T'=\gp(1+\bp)T$, and a fraction of $\gamma$-rays convert to $e^\pm$ pairs before exiting. The number of converting $\gamma$-rays per primary particle is
\beq
\label{eq:Mpm}
   \M_\pm=\frac{\bar{\sigma}_{\gamma\gamma}' U_\gamma' cT'}{m_ec^2} \, \frac{\dE_eT'}{m_ec^2}
    \approx \zeta\, \frac{cT}{R} \frac{\ssc'}{\sT} \,\ell^2\,\tausc,
\eeq
where 
\beq
   \ell\equiv \frac{\sT\E}{4\pi R^2 m_ec^2}\approx 66\,\frac{\E_{39}}{R_9^2},
\eeq
and $\bar{\sigma}_{\gamma\gamma}'=\zeta\,\sT$ is an effective cross section for $\gamma\gamma$ collisions. The numerical factor $\zeta$ may be found by integrating the $\gamma\gamma$ reaction rate over the spectrum of curvature emission. It depends on $\epsc'$ and varies around $\zeta\sim 10^{-2}$ at radii where $\epsc'>1$.

The $e^\pm$ created by $\gamma$-rays oscillate in the wave and emit secondary $\gamma$-rays just like the primary particles do. An avalanche of $e^\pm$ creation develops if $\M_\pm \gg 1$, boosting the plasma density by $e^{\M_\pm}$. One can see that $\M_\pm$ is large, even if one chooses a low density $n_0=\nco$,
\beq
   \M_\pm \sim \frac{10^8 \zeta}{R_9^7}\;\E_{39}^2\, T_{-3}\, \sigma_8^2 \, \N_{32} \gg 1,
\eeq
where $\sigma_8=\ssc'/10^8\sT$. The wave becomes immersed in the exponentially enriched $e^\pm$ plasma with $\tausc\gg 1$.

%#########################################################

\section{Discussion}

We have found $\tausc\gg 1$ for FRBs emitted deep inside the magnetosphere and attempting to escape through the outer magnetosphere. This result is obtained for a given strong radio burst, not including the feedback of energy losses on its luminosity $L$. In reality the burst luminosity must drop to satisfy $\tausc\simlt 1$, respecting energy conservation $\Esc\simlt\E$. Alternatively, the condition $\tausc(L,R)\sim 1$ may be used to define a minimum emission radius for a given $L$. Our results imply that observed FRBs cannot be released deep inside the magnetospheres of magnetars, at radii $R\ll 10^{10}\,$cm. This constraint is based on established laws of electrodynamics and should be robust. 

The constraint should hold for FRBs with any temporal structure, including double FRBs. The first burst will accelerate and displace the outer magnetosphere outside $R\sim 10^9\,$cm (Eq.~\ref{eq:bp}). Yet, this displacement creates no holes 
that could save the second burst from scattering --- it still must pass through the quasi-static magnetosphere just outside $R_0$ and then through the expanding layers at $R\simgt 10^9\,$cm. Note that the magnetospheric outflow launched by the scattering of the first FRB has a smooth radial profile: its speed increases from $\vp\approx 0$ in the inner layers to $\vp\sim c$ in the outer layers, and such smooth outflows do not develop holes. In addition, the outflow will get loaded with $e^\pm$ plasma created by $\gamma$-rays trailing the first FRB. Then, the perturbed magnetosphere will begin to relax toward a new equilibrium. Thus, the burst will fail to clear out an escape route and save subsequent FRBs from scattering.

Interestingly, lowering $L$ may not help FRBs escape, because $\ssc'\propto L^{-3/8}$ {\it grows} at lower $L$ (\Eq~\ref{eq:ssc}), and $\epsc'\propto L^{1/8}$ weakly changes. The main helping effect of lowering $L$ is the growth of $\Ra\propto L^{-1/4}$ (Fig.~\ref{fig:LR}). When $L\simlt 10^{36}\,$erg/s, $\Ra$ reaches $\sim 10^{10}\,$cm, the zone of strong particle acceleration $1<a_0<\omB/\omega$ disappears, and the wave scattering practically vanishes. The FRBs detected from SGR~1935+2154 are close to this transition.

We conclude that the scenario of a radio source confined in the inner magnetosphere is not satisfactory for FRBs in a broad range of observed luminosities. A remaining possibility for FRB emission by magnetars is a magnetic explosion --- the ejection of a large-scale electromagnetic pulse from the inner magnetosphere. The pulse has a size of $10^7$-$10^8$\,cm and is much more energetic than the observed GHz burst. \cite{Beloborodov17b,Beloborodov20} proposed that the pulse drives an ultra-relativistic shock in the magnetar wind and the shock emits coherent radio waves by the synchrotron maser mechanism. A version of this model (a shock striking slow ion ejecta) was developed by \cite{Metzger19}.  Alternatively, \cite{Lyubarsky14,Lyubarsky20} argued that the ejected pulse will freely propagate through the magnetar wind, like a vacuum electromagnetic wave (this can happen under certain conditions which will be discussed elsewhere). The field in the pulse is expected to have small-scale perturbations, and they may be released at a large radius as escaping radio waves \citep{Lyubarsky20}. 

Calculations of wave scattering in this {\it Letter} focused on the closed magnetosphere at $R<\RLC$. In a similar way, one can evaluate $\tausc$ in the wind at radii $R\gg\RLC$, where $n_0\sim \N/\RLC R^2$, $\Bbg\approx \mu/\RLC^2 R$, and scattering occurs in regime~I ($\omB'<\omega'$) with $\ssc'\approx a_0^2\sT$. With increasing radius, $\hbar\omsc'\approx a_0^3\hbar\omega'$ quickly falls below $m_ec^2$, so secondary $e^\pm$ creation becomes suppressed, and one finds 
\beq
 \tausc\sim \frac{a_0^2\sT \N}{\RLC R} \approx \frac{0.7\, L_{42}\N_{37}}{R_{10}^3\nu_9^2 (P/1\,{\rm s})} 
   \quad (R>10^{10}\,{\rm cm}).
\eeq
In particular, $\tausc$ is negligible if the observed FRBs are emitted by blast waves in the magnetar wind, which generate coherent radio emission at $R\sim 10^{13}-10^{14}\,$cm, although {\em induced} scattering could still be important in this model \citep{Beloborodov20}.

 \medskip

I thank Yuri Levin, Brian Metzger, and the referee for comments which helped improve the manuscript. This work is supported by NSF grant AST 2009453, Simons Foundation grant \#446228, and the Humboldt Foundation.

\bibliography{ms}

\end{document}